\documentclass[12pt]{article}
\usepackage{graphicx}
\usepackage{amsmath}
\usepackage{authblk}
\usepackage{epsfig} 
\usepackage{txfonts,ulem}
\usepackage{aas_macros}
\usepackage{color}
\usepackage{url}
\usepackage{natbib}
\usepackage{longtable}
\usepackage{times}
\usepackage{tikz}
\usepackage{hyperref}
\newcommand{\sepspeed}{$95 \pm 13$~m\,s$^{-1}$}
\newcommand{\leadspeed}{$121 \pm 22$~m\,s$^{-1}$}
\newcommand{\trailspeed}{$-70\pm 13$~m\,s$^{-1}$}

\title{SDO/HMI survey of emerging active regions for helioseismology
}
\author[1]{H. Schunker}
\author[2]{D.C.~Braun}
\author[1]{A.C.~Birch}
\author[1]{R.B.~Burston}
\author[1,3]{L.~Gizon}
\affil[1]{Max-Planck-Institut f\"{u}r Sonnensystemforschung, Justus-von-Liebig-Weg 3,  37077  G\"{o}ttingen, Germany}
\affil[2]{NorthWest Research Associates, 3380 Mitchell Ln, Boulder, CO 80301, USA}
\affil[3]{Georg-August-Universit\"{a}t G\"{o}ttingen, Institut f\"{u}r Astrophysik, Friedrich-Hund-Platz 1, 37077 G\"{o}ttingen	}

\begin{document}

\maketitle

\abstract
{
Observations from the Solar Dynamics Observatory (SDO) have the potential for allowing the helioseismic study of the formation of hundreds of active regions, which would enable us to perform statistical analyses.}
{
Our goal is to collate a uniform data set of emerging active regions observed by the SDO/HMI instrument suitable for helioseismic analysis, where each active region is centred on a $60^\circ \times 60^\circ$ area and can be observed up to seven~days before emergence.
}
{We restricted the sample to active regions that were visible in the continuum and emerged into quiet Sun largely avoiding pre-existing magnetic regions. As a reference data set we paired a control region (CR), with the same latitude and distance from central meridian, with each emerging active region (EAR). 
The  control regions do not have any strong emerging flux within 10$^\circ$ of the centre of the map. Each region was tracked at the Carrington rotation rate as it crossed the solar disk, within approximately $65^\circ$ from the central meridian  and up to seven days before, and seven days after, emergence. The mapped and tracked data, consisting of line-of-sight velocity, line-of-sight magnetic field, and intensity as observed by SDO/HMI, are stored in datacubes that are 410~minutes in duration and spaced 320~minutes apart.  We call this  data set, which is currently comprised of 105 emerging active regions observed between May 2010 and November 2012, the SDO Helioseismic Emerging Active Region (SDO/HEAR) survey. 
}
{
To demonstrate the utility of a data set of a large number of emerging active regions, we measure the relative east-west velocity of the leading and trailing polarities from the line-of-sight magnetogram maps during the first day after emergence. 
The latitudinally averaged line-of-sight magnetic field of all the EARs shows  that, on average, the leading (trailing) polarity moves in a prograde (retrograde) direction with a speed of \leadspeed (\trailspeed) relative to the Carrington rotation rate in the first day. However, relative to the differential rotation of the surface plasma, the east-west velocity is symmetric, with a mean of \sepspeed.
}
{
The SDO/HEAR data set will not only be useful for helioseismic studies, but will also be useful to study other features such as the surface magnetic field evolution of a large sample of EARs. We intend to extend this survey forwards in time to include more EARs observed by SDO/HMI.
}

\maketitle

\section{Introduction}\label{intro}

Surface magnetic activity is the most obvious characteristic of the internal solar dynamo. Understanding the formation, evolution and decay of active regions places tighter constraints on models of the solar dynamo, especially models of the generation of a poloidal magnetic field component from a rising toroidal field configuration.

The traditional view has been that flux tubes form deep in the convection zone and rise to pierce the surface forming the observed bipolar active regions \citep[see reviews, ][]{Fan2009,Charbonneau2014}.
More recently, it has been suggested that flux tubes may form in the bulk of the convection zone \citep{Nelson2014}. 
Further suggestions include scenarios where active regions are formed in the near-surface shear layer \citep[e.g.][]{Brandenburg2005}, or by strongly stratified hydromagnetic turbulence very close to the surface organising the surface magnetic field into coherent concentrations \citep[e.g.][]{Brandenburgetal2014}.

Three-dimensional numerical simulations of solar convection close to and including the solar surface \citep[e.g.][]{Cheungetal2010, Steinetal2011} provide critical insight into the physical mechanisms involved in different emergence scenarios.
For example, in some simulations \citep[e.g.][]{Cheungetal2010}, coherent bundles of flux are injected through the bottom boundary, while in others  \citep[e.g.][]{Steinetal2011} horizontal field placed at the bottom boundary are passively advected by the convective flows.
Helioseismology can potentially constrain and provide direct evidence for particular models but, as pointed out by \cite{Cheung2014}, the relationship between numerical modelling and helioseismology is a two-way street. In many cases the models themselves can provide validation of helioseismic methods, which are then applied to solar observations to constrain models of emergence. 
Efforts towards constraining the physics governing active region emergence in the Sun using simulations and helioseismology has already begun   \citep[][, in preparation]{Birchetal2016}.

Helioseismic case studies of individual active region emergences have been carried out for at least two decades \citep[e.g.][]{Braun1995,Chang1999,Jensenetal2001,Hartlep2011}. These case studies are useful and some have suggested strong pre-emergence features \citep[e.g.][]{Ilonidisetal2011}.  However, the different methodologies and goals make it difficult to form a consensus from disparate studies. Results can be apparently  contradictory even when comparing different studies of the same emerging regions \citep[e.g.][]{Ilonidisetal2011,Braun2012,Ilonidisetal2012,Braun2014}.
In addition, it is predicted that subsurface signals due to flux emergence may be small compared to the uncertainties in helioseismic inferences \citep{Birchetal2010}, which limits the utility of individual case studies for constraining models.

One way to look for weak signatures, as well as common properties of emergence, is to do a statistical 
analysis of many emerging active regions  \citep[e.g.][]{Kommetal2008,Kommetal2009,Birchetal2013}.
 \citet{Birchetal2010} estimated that by studying approximately 100 emerging active regions, helioseismology should be able to detect the predicted signature (a roughly 100~m\,s$^{-1}$ retrograde flow) of a buoyant rising flux tube below the surface before emergence. One  such study was undertaken using SOHO/MDI magnetic field observations and Global Oscillation Network Group (GONG) velocity observations \citep{Lekaetal2013,Birchetal2013,Barnesetal2014}, which we shall refer to as the `LBB survey'. \citet{Birchetal2013} detected weak converging (less than 15~ms$^{-1}$) flows in the day preceding emergence and mean travel-time perturbations apparently associated with surface magnetic features. The study was limited to about one day before emergence and was used to track regions spanning a $30^\circ \times 30^\circ$ area.
 
The data set presented in this paper was inspired by the LBB survey  \citep{Lekaetal2013,Birchetal2013,Barnesetal2014}.  Our data set, which we call the {\textit{Solar Dynamics Observatory Helioseismic Emerging Active Region (SDO/HEAR) Survey}} consists of 105 emerging active regions (EARs) observed by the Helioseismic and Magnetic Imager on board the Solar Dynamics Observatory (SDO/HMI) \citep{SDO2012} between  May 2010 (the start of science grade SDO/HMI observations) and November 2012 (when we began data selection and analysis). Each active region was observed up to seven days before and after emergence and covers a spatial region spanning $60^\circ \times 60^\circ$ surrounding the EARs, which allows helioseismic techniques to probe  substantially deeper below the surface than the LBB survey.

Section~\ref{sect:methear} and \ref{sect:methqs}  describe the selection criteria for the emerging active regions and of the control sample, 
respectively. Section~\ref{sect:datared} describes the data reduction of the magnetograms, Dopplergrams, and continuum images from SDO/HMI. 
Section~\ref{sect:indears} demonstrates the unique surface magnetic field features of active region  emergence, and Sect.~\ref{sect:aveears} describes the surface magnetic field pre- and  post-emergence characteristics of all of the EARs averaged together. In Sect.~\ref{sect:sepvel} we show an example study of the east-west velocity of bipoles in the first day after emergence using the SDO/HEAR data set.
We summarise our findings in Sect.~\ref{sect:summary} and highlight potential uses other than helioseismic analysis for this data set.

\section{Criteria for selecting emerging active regions}\label{sect:methear}
Our study is intended to both complement and extend the LBB survey.  We thus employ active  region selection criteria that are very similar to those employed in that work. The two  main properties of the emerging regions in this survey are essentially the same as those in the LBB survey: (1) they are visible in the continuum and (2) they emerge into relatively quiet (non-magnetic)  areas of the Sun not containing, nor in the immediate vicinity of, pre-existing active regions.
The motivation for selecting quiet emergence sites is to avoid or minimise the difficulty in distinguishing the contribution of any observed
signature related to pre-existing surface magnetic field from contributions due to emerging flux \citep[as summarised in ][]{Schunker2010}. 
An initial list of potential candidate emerging regions was first obtained from National Oceanic and Atmospheric Administration (NOAA) solar region reports for the time period 2010-2012. 
The reports used are the yearly summaries found at 
\url{http://www.ngdc.noaa.gov/stp/space-weather/solar-data/solar-features/sunspot-regions/usaf_mwl/}.
\begin{center}   
\begin{figure}
\includegraphics[width=0.8\textwidth]{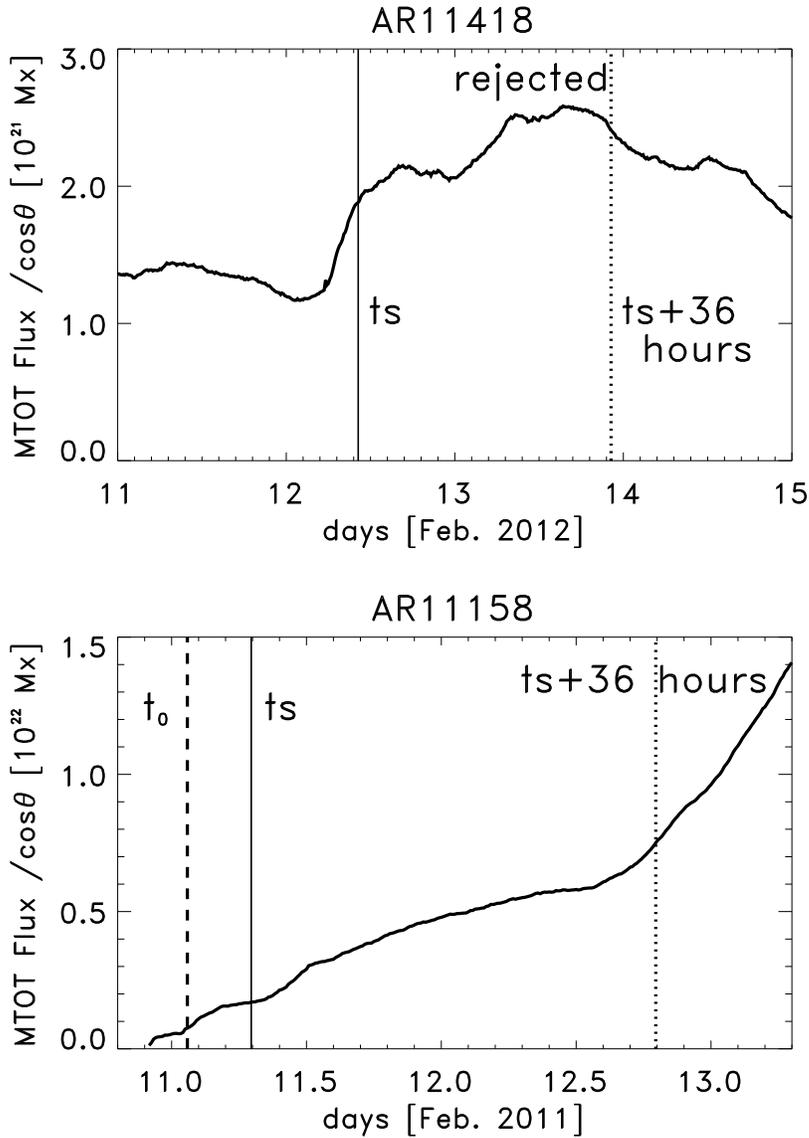} 
\caption{Corrected \texttt{MTOT} flux as a function of time for two example emerging active region candidates. The solid line represents the time, $t_{\rm s}$, when NOAA first recorded a  numbered sunspot region. The top panel shows an example  of an emerging sunspot group, AR~11418, which we rejected from our survey. While an increase of new flux is apparent near the time $t_{\rm s}$, it is obvious that considerable flux already exists within the SHARP boundary. The bottom panel shows an example, AR~11158, which we included in the survey.  The thick dashed line indicates the emergence time $t_0$ when the flux was 10\%  of  its peak value within the interval bounded by the solid and dotted lines, $t_{\rm s}$ and $t_{\rm s}+36$~hr, respectively.}
\label{fig:mtott}
\end{figure}
\end{center}

The NOAA records give the time, location (to the nearest 10$^\circ$ 
in heliographic coordinates), and a rough estimate of the area of
every sunspot group assigned a number. For each numbered region we identified the time, which we call here $t_{\rm s}$, of its first observation. For our survey, we  selected only NOAA numbered regions that reached a total area  of at least 10~$\mu{\rm H}$ (micro hemispheres, $1~\mu{\rm H} \approx 3$~Mm$^2$) to be consistent with the previous LBB Survey. The NOAA tabulated areas, which are estimated by eye, are in increments of 10~$\mu{\rm H}$. It is possible for a region to be assigned a value of 0, but all NOAA numbered regions during the interval of our survey reached at least 10~$\mu{\rm H}$ during their disk passage. 
At time $t_{\rm s}$ the region must be located within $50^\circ$ of the central meridian. 
This restriction removed from further consideration regions that emerged on the far side or near the limb of the Sun.
This same restriction kept regions that could be observed out to heliocentric angles of $60-65^\circ$, thereby limiting the  centre-to-limb effects, such as foreshortening, that may compromise local-helioseismic analysis.
The two  selection criteria (area and central meridian distance at $t_{\rm s}$) resulted in 239 candidate emerging active regions over the 31~month interval. 

To define emergence times and map centres,  and to select active regions that emerge into quiet Sun,  we used characteristics of the magnetic flux measured within  Space-weather HMI Active Region Patches (SHARPs) as assessed and stored at the Joint Science Operations Center (JSOC) headquartered at Stanford University \citep{Bogart2007}. In particular we used the data series labelled
\texttt{hmi.sharp\_720s,} which identifies and characterises active regions on  the surface of the Sun \citep{SHARP2014}.
The patches are regions in the HMI line-of-sight magnetograms that represent enduring,  coherent magnetic structures on the size scale of a solar active region. They are labelled by an assigned \texttt{HARP} number and may cover more than one NOAA numbered region. 

Consistent with the LBB survey, we define the emergence time $t_0$ as the time when  the absolute flux, corrected for line-of-sight projection, reaches 10\% of its maximum value over a 36-hour interval following the first appearance of the sunspot (or group) in the NOAA record (time $t_{\rm s}$ in Fig.~\ref{fig:mtott}). This time often coincides with a rapid increase  of the kurtosis (fourth moment) of the spatial distribution of the line-of-sight magnetic field \citep{Lekaetal2013}.  In general, the emergence time $t_0$ occurs before the time of the first NOAA observation $t_{\rm s}$ (e.g. see bottom panel of Fig.~\ref{fig:mtott}).

The absolute line-of-sight flux within a SHARP  is given by the \texttt{MTOT} keyword in  the \texttt{hmi.sharp\_720s} series. 
We (approximately) corrected \texttt{MTOT} for the line-of-sight  projection by dividing it by $\cos \theta$, where $\theta$ is the angular distance to disk centre (this assumes that the magnetic field is radial at the solar surface). By visually inspecting plots of this corrected flux with time, we were able to identify and reject many regions that emerged into or close to the existing flux. 
In particular, regions kept in the survey showed values of \texttt{MTOT} that rise monotonically and sharply from a low, constant flux, whereas the flux of rejected regions showed a more complicated variation with time.
Figure~\ref{fig:mtott} shows examples of a rejected candidate, AR~11418, and a selected candidate, AR~11158.  After inspecting flux plots for all 186 HARP/NOAA regions, we rejected 81.  This left 105 emerging active regions for our survey.  For these regions we recorded the emergence time $t_0$, as defined above, and the heliographic coordinates of the region (see Table~\ref{tab:ears}). 
The coordinates were defined by the flux-weighted centre of the line-of-sight magnetic field at the emergence time and were derived from the Stonyhurst coordinates specified by the SHARP keywords \texttt{LAT\_FWT} and  \texttt{LON\_FWT}.

\begin{center}
\begin{figure}
\includegraphics[width=0.8\textwidth]{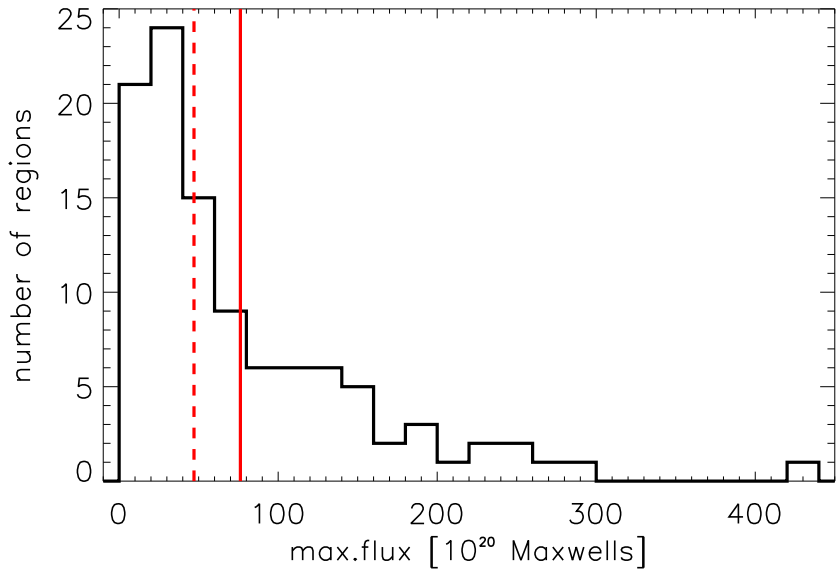} 
\caption{Number of EARs as a function of maximum total unsigned  flux. The 
solid red vertical line indicates the mean and the dashed line indicates the 
median value. Active regions indicated with asterisks in Table~\ref{tab:ears} are 
those with values greater than the median.
 }
\label{fig:fluxrate}
\end{figure}
\end{center}

In the LBB survey the emergence time was defined within a cadence of 96~minutes, but in our survey using HMI observations we have defined it within a cadence of 12~minutes, which is the cadence of the \texttt{MTOT} values.
Figure~\ref{fig:fluxrate} shows the distribution of maximum total unsigned flux of the emerging active regions, given by the keyword  in the 
\texttt{hmi.sharp\_720s} data series \texttt{USFLUX}$=\sum_{i} |B_z| \mathrm{d}A$, where $B_z$ is the  radial component of the vector magnetic field and $A$ is the area enclosed by the bounding curve. Table~\ref{tab:ears}  provides the list of all EARs along with their emergence times and locations. 

Once the data reduction was carried out (see Sect.~\ref{sect:datared}), we assigned a number, $P$, to indicate the amount of pre-emergence flux (see Table~\ref{tab:ears}) based on a visual inspection of the mapped magnetograms. This subjective factor reflects the presence of a nearby magnetic field (from plage or active regions), which could not otherwise be identified from the flux as a function of time described above, and is intended to indicate the suitability of the active region for unambiguous helioseismic studies. A $P$-factor of 0 represents an emergence into a very quiet region (21 regions); a $P$ factor of 1 or 2 indicates emergence into increasing amounts of magnetic field nearby (but not directly at) the subsequent emergence location (59 regions); and a $P$-factor of 3 or higher indicates the region may be compromised by pre-existing field at the emergence time and location (25 regions). The values of 0, 1, 2 are assigned according to whether the nearby field looks to the eye to be less than, typical of, or more than the average. We note that even though regions with a $P$ factor higher than 2 may not be suitable for an unambiguous helioseismic analysis, the data set in its entirety forms a consistently selected set that may have value for a variety of purposes, including to understand more complicated emergence processes.

\section{Criteria for selecting control regions}\label{sect:methqs}
\begin{center}
\begin{figure}
\includegraphics[width=0.8\textwidth]{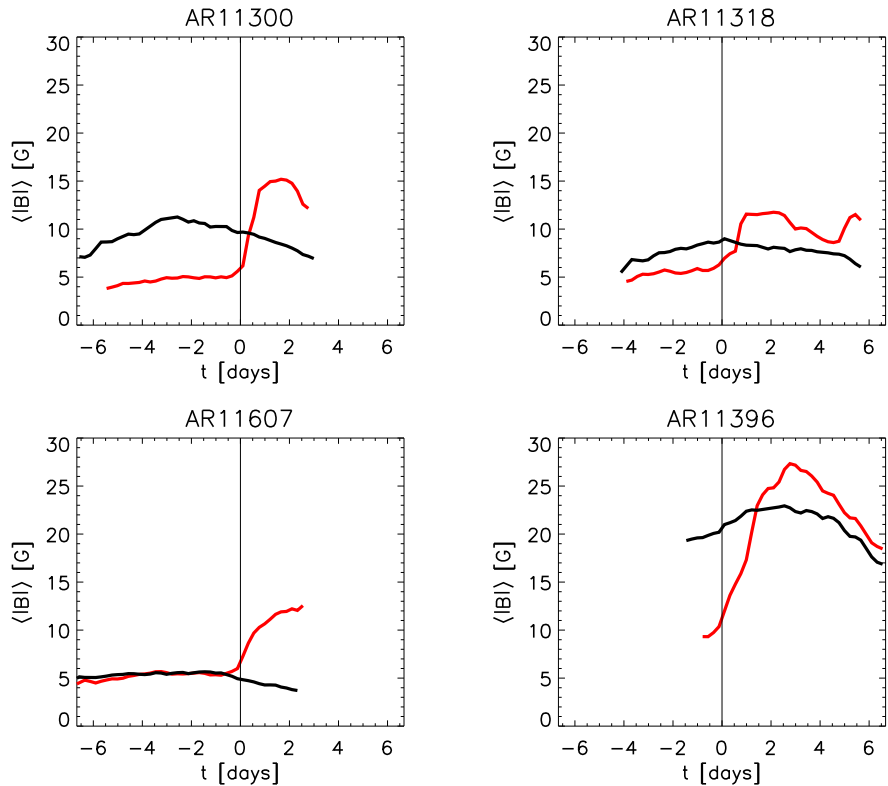}
\caption{Examples of the mean  absolute line-of-sight magnetic field strength 
in the central $10^\circ$ for a sample of EARs (red) and corresponding CRs 
(black) as a function of time relative to the emergence time. }
\label{fig:earscrs}
\end{figure}
\end{center}

Crucial to correctly interpreting pre-emergence signatures, or the variation with time of any observed variable obtained from our survey, is to also have a control  sample that forms a basis for comparison. Such comparisons can detect the presence of systematic effects, which may, for example, depend on the position of the observations on the solar disk or on time. Ideally, the control sample should have similar if not identical distributions in their properties, such as the phase of the solar cycle or their latitude and central meridian distance. This goal was achieved by producing a control sample of quiet regions (hereafter simply control regions or CR) that were paired with the EARs, such that each pair crossed the disk at the same latitude but were separated in time by no more than two solar rotation periods.  
We automated the initial selection of CRs as follows:
\begin{itemize}
\item For each EAR we searched for a corresponding CR by adjusting a variable mock-emergence time $t_{\rm c}$ at which the Stonyhurst coordinates of the CR was the same as the EAR at its (real) emergence time $t_0$.
\item We began searching two days forwards in time $(t_{\rm c} = t_0 + 2$~d),  incrementing $t_c$ in three day intervals up to 60 days. 
\item If there were no numbered SHARP regions within an $18^\circ$ radius of  the central map location and the absolute difference in $B_0$ angle was less than $8^\circ$,  then this was considered a potential control region.  These numbers were a result of trial and error to ensure that a potential CR  could be found for all EARs. If none were found forwards in time, then the search continued in the same way backwards in time.
\end{itemize}
We then tracked and mapped magnetogram datacubes (see Sect.~\ref{sect:datared}) and computed the average  magnetograms for the chosen regions. We plotted the mean absolute magnetic  field in the central $10^\circ$ of the map and visually inspected the  maps to detect any signs of nearby emergence, as in Fig.~\ref{fig:earscrs}. 

\begin{center}
\begin{figure}
\includegraphics[width=0.8\textwidth]{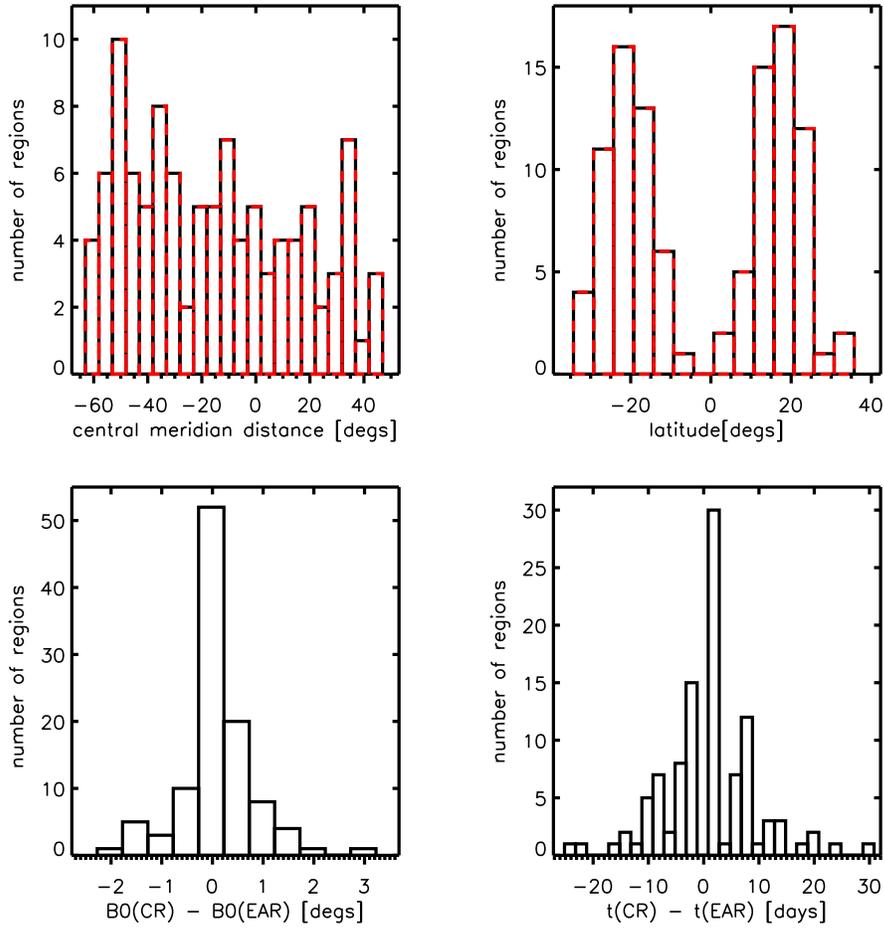}
\caption{Histogram of the control regions (solid black) and emerging active 
regions (red dashed) as a function of central meridian distance (top left), 
latitude (top right), the difference in $B0$-angle (bottom left), and the  
difference between emergence times (bottom right).  }
\label{fig:stats}
\end{figure}
\end{center}

If the CR did not show an increase in flux similar to an emergence and the absolute difference between the average pre-emergence flux of the CR and EAR was less than $\approx 10$~G then this region was selected  as the paired control region.  We found that 34 regions out of the 105 did not have a sufficiently quiet CR within 60 days of the emergence time $t_0$. For these regions we searched further forwards or backwards in time until an acceptable CR was identified. Figure~\ref{fig:stats} shows the distributions of the EAR-CR pairs.

This process is in contrast to the LBB survey, where control regions  were specifically selected to be regions with absolute magnetic field values  of consistently less than 1~kG for the three days before emergence. 
Table~\ref{tab:ears} provides the list of all EARs and corresponding CRs in the  data set. 

\section{Data reduction}\label{sect:datared}
\begin{center}
\begin{figure}
\includegraphics[width=0.8\textwidth]{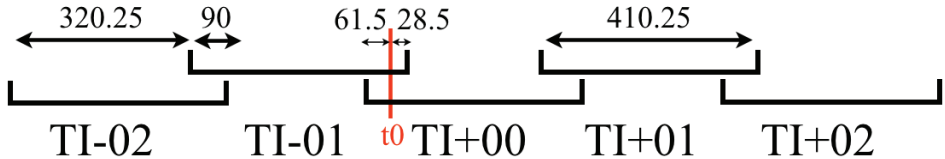}
\caption{
Sketch of the distribution of datacubes in time. The rectangular sections 
indicate the extent of single datacubes, labelled by the time interval number 
``TI''. The numbers across the top are in minutes.
Intervals start every 320~minutes,  are 410.25~minutes long, and overlap by 
90~minutes.  We show only the datacubes close to the emergence time, $t_0$, 
indicated in red. In general, the EAR and CR extend both further forwards and 
backwards in time.
}
\label{fig:tstep}
\end{figure}
\end{center}

The primary data analysis consisted of tracking and remapping  SDO/HMI line-of-sight magnetogram, Dopplergram, and 
intensity observations, for up to a two-week duration, of each EAR and CR in the survey. 
The data analysis and storage was carried out within the German Data Centre for SDO  \citep{Burstonetal2008,Saidi2010}, housed at the Max Planck Insitute for Solar System research in G\"ottingen, Germany.
We tracked the EARs and CRs (Table~\ref{tab:ears}) at the Carrington rotation  rate over intervals of 6.825~hours (547 frames with a cadence of 45~seconds),  spaced 5.3375~hours (320.25~minutes, 427 frames) apart. 
Thus, there is a 1.5~hour (120 frame) overlap (see Fig.~\ref{fig:tstep}). 
The tracking rate and spacing of the datacubes was taken directly from the LBB survey. 
The 15 second difference in length of the datacubes is due to the 45 second cadence of the SDO/HMI observations and the 60 second cadence of the GONG observations.
The temporal overlap between successive datacubes (90 min) is slightly longer than used in the LBB survey (64~min).
 
The emergence time occurs 28.5~minutes (39 frames) before the end of the  datacube labelled with time interval \texttt{TI=-01} and 1 hour and 45 seconds (81 frames) after the beginning of the datacube labelled \texttt{TI=+00}.
This is designed to align the time intervals with the emergence time $t_0$ in a nearly identical manner to the LBB survey.

\begin{center}
\begin{figure}
\includegraphics[width=0.8\textwidth]{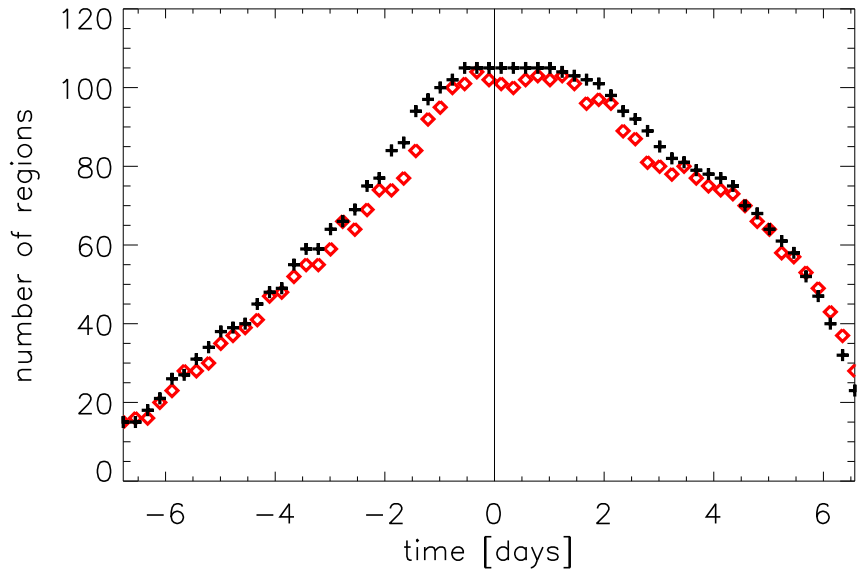} 
\caption{Number of EARs (red diamonds) and CRs (black crosses) of the line-of-sight magnetogram datacubes as at each time interval. }
\label{fig:num}
\end{figure}
\end{center}

At each 45 second interval we Postel projected full-disk SDO/HMI observations onto  $60^\circ \times 60^\circ$ maps. The projection is made to a $512 \times 512$~pixel grid with a pixel size of $1.4$~Mm, and is centred at the specified flux-weighted centroid position described in Section~\ref{sect:methear} and listed for each EAR and CR in Table~\ref{tab:ears}.
The pixel size is about four times larger than the raw HMI images.
The size of the maps is four times larger, spanning twice the length on each side, than the LBB survey allowing more flexibility
for helioseismic analysis.
The large-scale rotational signal and orbital velocity of the satellite in the Dopplergrams are removed by fitting and subtracting a plane to each map.
The number of emerging active regions with data at each time interval relative to emergence is shown in Fig.~\ref{fig:num}. The duty cycles for individual datacubes are typically greater than $95$\%.

\begin{center}
\begin{figure}
\hspace{-0.5cm}
\includegraphics[width=0.8\textwidth]{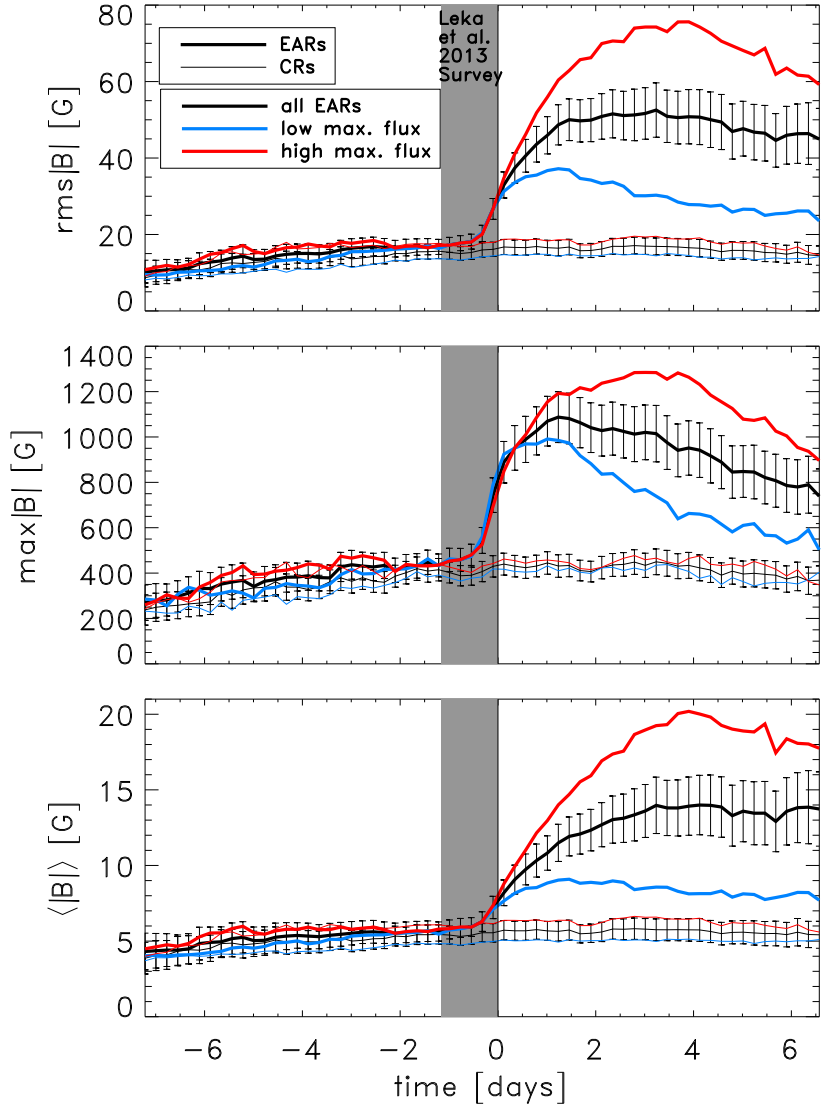} 
\caption{Mean (bottom panel), maximum (middle panel), and rms (top panel) of the absolute line-of-sight magnetic field in the central 10$^\circ$ radius averaged over all emerging active regions (thick black) at each time interval. 
The blue curve shows the values for only those active regions that reach a maximum flux that is lower than the median maximum flux value, and the red curve shows the values for the emerging active regions that reach a higher value.
The corresponding values for the control regions are shown by the thin curves.
The shaded grey area indicates the time span covered by the previous LBB survey.
The uncertainties are the rms of the quantity. For example, in the top panel, $\sigma_{B_\mathrm{rms}}=\sqrt{{\sum B_\mathrm{rms}^2}/{N}}$, where $N$ is the number of active regions at time interval, \texttt{TI}. 
}
\label{fig:meanb}
\end{figure}
\end{center}

To describe the spatial distribution of the magnetic field, we show the spatial average, maximum value, and rms of the absolute line-of-sight magnetic field averaged over the 105 regions as a function of time in Fig.~\ref{fig:meanb} over a central 10$^\circ$ disk.
Also shown are the same values averaged over the two halves of the data set separated by the median value of their maximum flux ($47\times10^{20}$ Maxwells).  
The CRs paired with EARs with a lower maximum flux have weaker magnetic field properties than those CRs paired with high maximum flux EARs (see the difference between the thin blue and red curves in Fig.~\ref{fig:meanb}). 
This is likely because of a selection effect: the paired CRs for the smaller flux regions were selected to predominantly have a lower mean magnetic field than the CRs paired with the EARs.

\section{Individual emerging active regions: Examples}\label{sect:indears}
\begin{center}
\begin{figure}
\hspace{-1cm}
\includegraphics[width=0.8\textwidth]{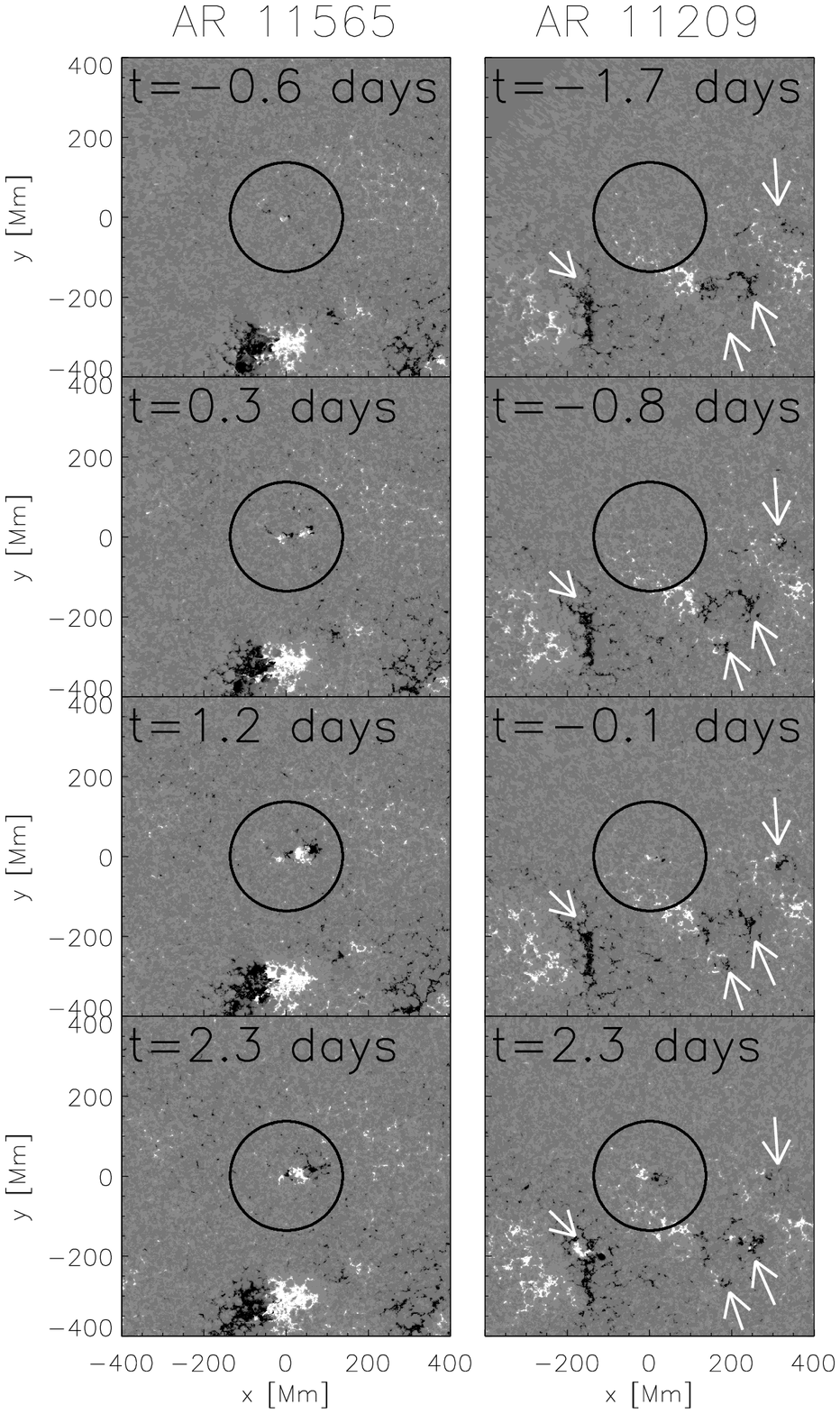} 
\caption{Evolution of the line-of-sight magnetic field of two emerging active  regions, AR~11565 (left column) and AR~11209 (right column). AR~11565 shows a double emergence and AR~11209 is a smaller active region 
with many surrounding small emergences pointed to by the arrows. The black circle indicates a radius of $10^\circ$ from the centre of the  emerging active region. The greyscale is saturated at $\pm200$~G.
}
\label{fig:indears}
\end{figure}
\end{center}

Each active region emergence is unique. Two examples are shown in Fig.~\ref{fig:indears}.
Active region (AR) 11565 is an example of a `double emergence' where, just after the emergence of the main dipole, a smaller dipole emerges to the east and promptly decays. An example of a smaller active region emergence is AR~11209; here there are many surrounding smaller instances of emergence that occur almost simultaneously. These two active regions have different surrounding magnetic field conditions. AR~11565 has a strong active region to the south, whereas AR~11209 has more diffuse field in the surrounding region. 
These two active regions are only examples of the different types of emergence seen in our survey. We emphasise that studies of subsets and case studies of the SDO/HEAR Survey are important to further characterise the emergence process, for example studying the differences in emergence properties from weak and strong flux active regions.

\section{Surface properties of the average emerging active region}\label{sect:aveears}

To follow the spatial line-of-sight magnetic field evolution of an average emerging active region, we temporally averaged each magnetogram datacube to get the 410-minute  average magnetic field map for each time interval.
We aligned the temporally averaged magnetograms of each active region by shifting the magnetic field map to a common emergence location determined by the following method.
We defined the emergence location of each active region by taking the absolute  difference of the  averaged magnetogram map at time $t=-19$~hours 
(\texttt{TI=-04}) and $t=13$~hours (\texttt{TI=+02}).   We then determined the  centroid of the pixels within $3.5^\circ$ of the centre of the map and where the absolute difference magnetogram is larger than 30\% of its maximum value.
Once shifted to the emergence centre,  we computed the mean magnetic field map of the ensemble of EARs for each time interval. 

\begin{center}
\begin{figure}
\includegraphics[width=0.8\textwidth]{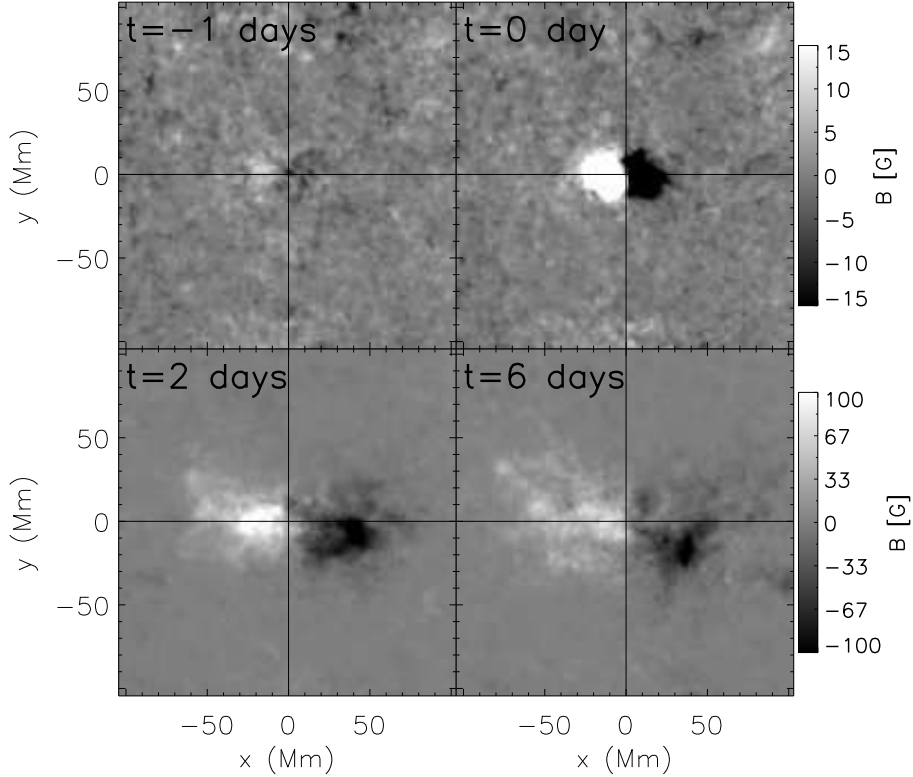} 
\caption{Average line-of-sight magnetic field maps for all EARs at selected time intervals. The southern hemisphere active regions have had their polarity reversed and have been flipped in the $y$ direction to account for Hale's law.}
\label{fig:aveear}
\end{figure}
\end{center}

Hale's polarity law states that most regions of strong magnetic field are grouped in pairs of opposite polarity regions, which are roughly aligned east-west with the leading polarity closer to the equator (Joy's law). 
Therefore, to account for Hale's law when averaging EARs in the north and south hemispheres, we reversed the magnetic field polarity of the regions in the southern hemisphere so that the leading polarity is always negative and flipped the maps in the latitudinal direction to account for Joy's law (i.e. so that the $+y$ direction points away from the equator). 
Figure~\ref{fig:aveear} shows four time intervals of the ensemble-averaged,  line-of-sight magnetogram maps. 
In the average emergence the polarities are  closer together, more compact and have a weaker magnetic flux at the emergence time compared to later times. The magnetic field strength of both polarities rapidly increases over one day and becomes more diffuse with time. Joy's law is clearly observed from two days after emergence, with the leading polarity closer to the equator than the following polarity. At $t=6$~days, the averaged line-of-sight magnetogram map is dominated by emerging regions with a larger maximum flux since they tend to have longer lifetimes, as can be seen in Fig.~\ref{fig:meanb}.

\section{Active region polarity separations}\label{sect:sepvel}
It has been suggested that the geometry and rise speed of rising flux tubes can be inferred from the motions of bipolar sunspot groups \citep[see][, and references therein]{vDGGreen2015}.
\cite{vDGP1990} measured the horizontal velocity of the leading and trailing polarities   of 95 relatively simple bipolar sunspot groups and found that the leading polarities move westwards at a speed between $100-700$~m~s$^{-1}$ and that the trailing polarities moved eastwards at $<100$~m~s$^{-1}$ relative to Carrington coordinates. \cite{Weberetal2013} also measured the rotation rates for the leading and trailing polarities in active regions.  They found that the leading polarity tends to rotate faster than the trailing polarity \citep[Fig.~9 of][]{Weberetal2013}, but do not specifically address a theoretical reason for this.
\cite{McClintockNorton2016} also measured the separation of the umbrae of $\approx 200$ sunspot groups observed in SDO/HMI intensity continuum and found that larger umbrae separate significantly more than smaller umbrae. 

\begin{center}
\begin{figure}
\includegraphics[width=0.8\textwidth]{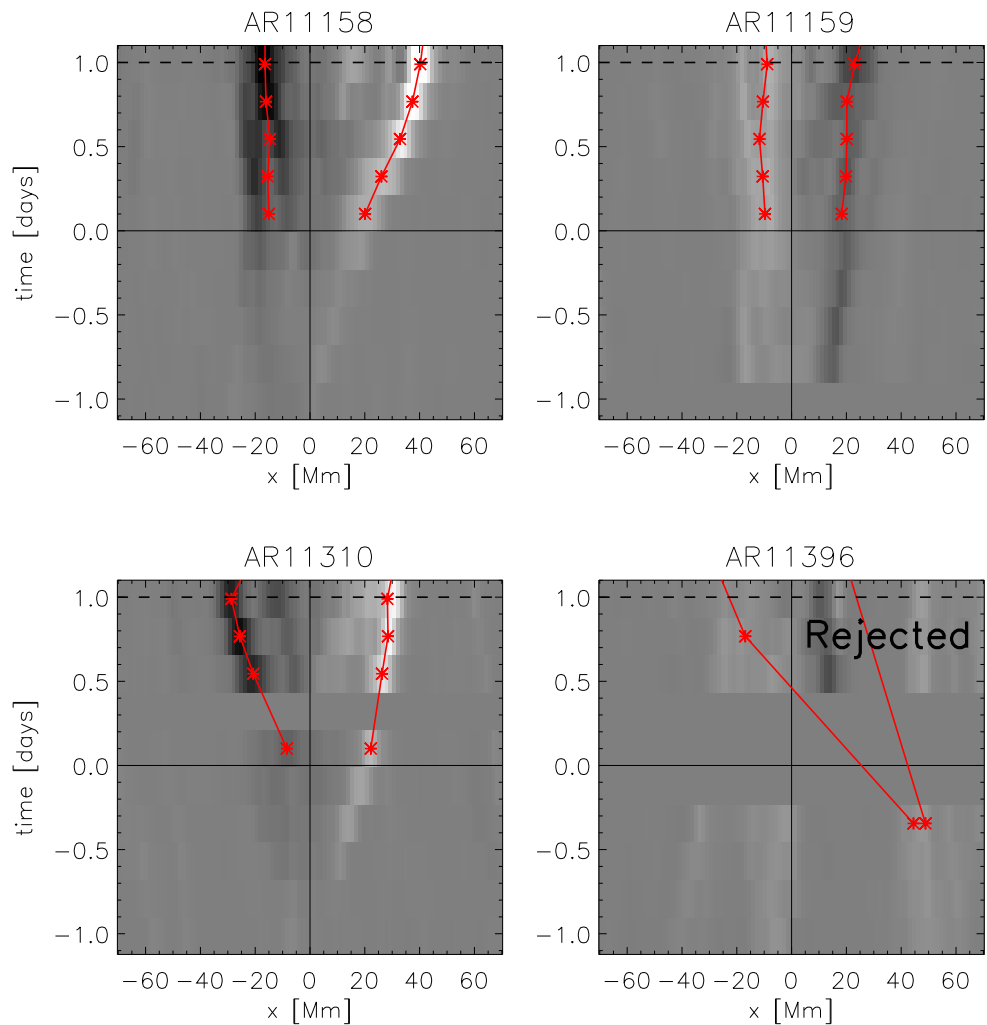}
\caption{Examples of the temporally averaged line-of-sight magnetic field averaged over the central 15~Mm in the $y$ direction for selected emerging active regions. The greyscale is saturated at $\pm 500$~G. The red crosses show the centre of the fitted Gaussian to the leading (positive $x$ direction) and trailing (negative $x$ direction) polarities.  
}
\label{fig:exlongtime}
\end{figure}
\end{center}

At each time interval we averaged the line-of-sight magnetic field across the central $15$~Mm in the north-south direction (Fig.~\ref{fig:exlongtime}). If the maximum absolute value of the magnetic field is greater than the mean of the absolute value by at least 20~G, then we performed a non-linear least-squares fit of a Gaussian  to the magnetic field as a function of longitude out to $6^\circ$ (70~Mm) from the centre of the active region eastwards (westwards) for the trailing (leading) polarity. 
From this, the east-west location of the Gaussian peak was defined as the location of the polarity. The motion of the peak in time gives us the east-west velocity of the polarities relative to the Carrington rotation rate (the rate at which the data cubes were tracked). We computed the east-west velocity of each polarity in each emerging active region.

We then computed the mean of the leading and trailing velocities in the first day by averaging the separation speeds for time intervals TI+00 to TI+04 for each EAR.
In some cases it was not possible to measure the separation speed reliably, for example AR~11396 in Fig.~\ref{fig:exlongtime}. In this case there was data missing near the emergence time and the region was small and weak, and thus our analysis method was not effective. The footer to Table~\ref{tab:ears} lists the regions for which our analysis method was not suitable and a velocity was not measured.

\begin{center}
\begin{figure}
\includegraphics[width=0.8\textwidth]{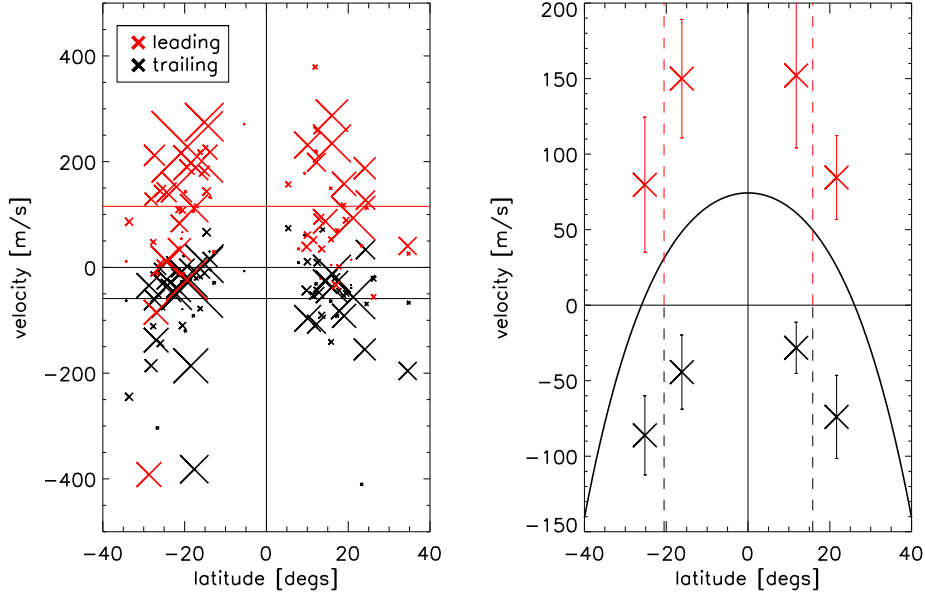} 
\caption{Left: the mean east-west velocity relative to the Carrington rotation rate of the leading (red crosses) and trailing (black crosses) polarities over the first day after emergence for each active region.  
The size of the symbols represents the size of the active region (AR~11158 is the largest). The scatter is large; this emphasises the uniqueness of each active region emergence. 
The mean velocity of the leading polarity in the first day after emergence is \leadspeed and the trailing polarity is \trailspeed.
Right: the average velocities in bins of polewards and equatorwards latitudes divided by the median latitude (dashed vertical lines) of the EARs.
The black curve shows the differential velocity of the surface plasma relative to the Carrington rotation rate. The uncertainties are given by the rms of the velocities in each bin, divided by the square root of the number of EARs in the bin. }
\label{fig:avesepvel}
\end{figure}
\end{center}

We emphasise again that the measured velocities differ greatly between active regions. Figure~\ref{fig:avesepvel} (left) shows the large scatter in the  velocity of the EARs in the first day after emergence.
To measure the latitudinal dependence of the east-west separation speed of the polarities, we averaged the velocities into bins of equatorwards and polewards active regions in each hemisphere (Fig.~\ref{fig:avesepvel}, right). The average separation speed was not weighted by independent uncertainties for each active region.
         
The mean east-west velocity of the leading polarity in the first day after emergence is $122$~m\,s$^{-1}$ and the trailing polarity is $-70$~m\,s$^{-1}$, relative to the Carrington rotation rate. However, relative to the surface plasma differential rotation \citep{SnodgrassUlrich1990} the velocities are  symmetric to within the uncertainties with a mean east-west velocity of each polarity relative to the emergence location of \sepspeed ; the separation speed between the leading and trailing polarities is twice this number.

Fitting a single Gaussian to the magnetic field in the leading and trailing polarity regions was an empirical choice that works well for typical emergence cases. 
This method is not suitable for double emergences and complicated flux evolution, which usually only appears later than two days after emergence. For this reason we only compute the average east-west velocity for the first day in Fig.~\ref{fig:avesepvel}. Analysing the velocity of the polarities further forwards in time would require a more detailed analysis, and for later times would be biased towards the longer living, larger active regions.

We repeated the same analysis on the 45~s cadence datacubes, which resulted in a separation speed of $83 \pm 13$~m\,s$^{-1}$.
We also experimented with averaging different widths in the north-south direction, since we may not be capturing the full development of the larger active regions. If we take only a cut in the north-south direction at $y=0$ through the centre of the temporally averaged magnetic field maps at each time interval, the separation velocity is $81 \pm 14$~m\,s$^{-1}$, and if we increase the width of the north-south averaging to 28~Mm the measured separation speed is $93 \pm 15$~m\,s$^{-1}$ . 
For the nominal case, we also computed the cross-correlation between consecutive times of the north-south averaged magnetic field. By fitting a quadratic to the five points centred on the maximum of the cross-correlation, we identified the shift in the polarity between the time steps.
This gave us a mean separation speed of $99 \pm 11$~m\,s$^{-1}$. 
In all cases the symmetry of the separation speed as a function of latitude about the differential rotation profile persists.

\section{Summary}\label{sect:summary}
We designed this data set  specifically to measure the subsurface signal of emerging active regions using helioseismology,  prompted by the previous LBB survey. The active regions in the SDO/HEAR survey are tracked up to seven days before and after emergence and the maps cover a larger area allowing helioseismic analysis to probe substantially deeper below the surface. Our selection of the paired control regions is different than in the LBB survey in that we do not enforce an upper limit on the magnetic flux.
The broad characteristics of these active regions are that they are all visible in the continuum and do not emerge into existing strong magnetic field.

This data set is also useful for studying the surface magnetic field properties, as shown in this paper.
As one example, we found that, on average, the speed of the leading polarity of all EARs is symmetric to the trailing polarity relative to the surface plasma differential rotation rate, which may be important for theories of active region formation. The mean east-west separation speed of each of the polarities relative to the emergence location first day is \sepspeed. The value quoted here is computed from first averaging the line-of-sight magnetic field averaged over the central 15~Mm in the north-south direction and then fitting a Gaussian to each polarity as a function of time. Other measurement techniques give different values of the same order.

In addition to studying active region emergence, the control regions described here also provide an opportunity to  study plage regions and perhaps weaker magnetic bipoles that do not develop into strong active regions.

In the future, it will be useful to expand the catalogue to include additional emerging active regions observed by SDO/HMI since the end of 2012, which at the present are estimated to at least double the numbers in this survey.
The SDO/HMI observations contain  a wealth of unexplored information.
The SDO/HEAR Survey data set we have collated is a preprocessed, standardised set ideal for statistical analyses of active region formation and evolution with helioseismology.

\section*{Acknowledgments}
The authors would like to thank the anonymous referee for his/her constructive comments.
Data courtesy of NASA/SDO and the HMI science team.
The German Data Center for SDO, funded by the German Aerospace Center (DLR), provided the IT infrastructure for this project.   This work utilised the Pegasus workflow management system.
The participation of DCB in this project is supported by the NASA Heliophysics Division through contracts NNH12CF23C and NNH12CF68C, and by the Solar Terrestrial program of the National Science Foundation through grant AGS-1127327.
LG acknowledges support from the Center for Space Science at the NYU Abu Dhabi Institute, UAE, under grant G1502.

\bibliographystyle{aa} 
\bibliography{ears_arxiv} 

\begin{thebibliography}{34}
\expandafter\ifx\csname natexlab\endcsname\relax\def\natexlab#1{#1}\fi

\bibitem[{{Barnes} {et~al.}(2014){Barnes}, {Birch}, {Leka}, \&
  {Braun}}]{Barnesetal2014}
{Barnes}, G., {Birch}, A.~C., {Leka}, K.~D., \& {Braun}, D.~C. 2014, \apj, 786,
  19

\bibitem[{{Birch} {et~al.}(2010){Birch}, {Braun}, \& {Fan}}]{Birchetal2010}
{Birch}, A.~C., {Braun}, D.~C., \& {Fan}, Y. 2010, \apjl, 723, L190

\bibitem[{{Birch} {et~al.}(2013){Birch}, {Braun}, {Leka}, {Barnes}, \&
  {Javornik}}]{Birchetal2013}
{Birch}, A.~C., {Braun}, D.~C., {Leka}, K.~D., {Barnes}, G., \& {Javornik}, B.
  2013, \apj, 762, 131

\bibitem[{{Birch} {et~al.}(2016){Birch}, {Schunker}, {Braun}, {Cameron},
  {Gizon}, {Loeptien}, \& {Rempel}}]{Birchetal2016}
{Birch}, A.~C., {Schunker}, H.~S., {Braun}, D.~C., {et~al.} 2016, Science
  Advances

\bibitem[{{Bobra} {et~al.}(2014){Bobra}, {Sun}, {Hoeksema}, {Turmon}, {Liu},
  {Hayashi}, {Barnes}, \& {Leka}}]{SHARP2014}
{Bobra}, M.~G., {Sun}, X., {Hoeksema}, J.~T., {et~al.} 2014, \solphys, 289,
  3549

\bibitem[{{Bogart}(2007)}]{Bogart2007}
{Bogart}, R.~S. 2007, Astronomische Nachrichten, 328, 352

\bibitem[{{Brandenburg}(2005)}]{Brandenburg2005}
{Brandenburg}, A. 2005, \apj, 625, 539

\bibitem[{{Brandenburg} {et~al.}(2014){Brandenburg}, {Gressel}, {Jabbari},
  {Kleeorin}, \& {Rogachevskii}}]{Brandenburgetal2014}
{Brandenburg}, A., {Gressel}, O., {Jabbari}, S., {Kleeorin}, N., \&
  {Rogachevskii}, I. 2014, \aap, 562, A53

\bibitem[{{Braun}(1995)}]{Braun1995}
{Braun}, D.~C. 1995, in GONG 1994. Helio- and Astro-Seismology from the Earth
  and Space, ed. R.~K. {Ulrich}, E.~J. {Rhodes}, Jr., \& W.~{Dappen}, Vol.~76
  (San Francisco: Astron. Soc. Pacific), 250

\bibitem[{{Braun}(2012)}]{Braun2012}
{Braun}, D.~C. 2012, Science, 336, 296

\bibitem[{{Braun}(2014)}]{Braun2014}
{Braun}, D.~C. 2014, \solphys, 289, 459

\bibitem[{{Burston} {et~al.}(2008){Burston}, {Gizon}, {Saidi}, \&
  {Solanki}}]{Burstonetal2008}
{Burston}, R., {Gizon}, L., {Saidi}, Y., \& {Solanki}, S.~K. 2008,
  Communications in Asteroseismology, 157, 293

\bibitem[{{Chang} {et~al.}(1999){Chang}, {Chou}, \& {Sun}}]{Chang1999}
{Chang}, H.-K., {Chou}, D.-Y., \& {Sun}, M.-T. 1999, \apjl, 526, L53

\bibitem[{{Charbonneau}(2014)}]{Charbonneau2014}
{Charbonneau}, P. 2014, \araa, 52, 251

\bibitem[{{Cheung} \& {Isobe}(2014)}]{Cheung2014}
{Cheung}, M.~C.~M. \& {Isobe}, H. 2014, Living Reviews in Solar Physics, 11, 3

\bibitem[{{Cheung} {et~al.}(2010){Cheung}, {Rempel}, {Title}, \&
  {Sch{\"u}ssler}}]{Cheungetal2010}
{Cheung}, M.~C.~M., {Rempel}, M., {Title}, A.~M., \& {Sch{\"u}ssler}, M. 2010,
  \apj, 720, 233

\bibitem[{{Fan}(2009)}]{Fan2009}
{Fan}, Y. 2009, Living Reviews in Solar Physics, 6, 4

\bibitem[{{Hartlep} {et~al.}(2011){Hartlep}, {Kosovichev}, {Zhao}, \&
  {Mansour}}]{Hartlep2011}
{Hartlep}, T., {Kosovichev}, A.~G., {Zhao}, J., \& {Mansour}, N.~N. 2011,
  \solphys, 268, 321

\bibitem[{{Ilonidis} {et~al.}(2011){Ilonidis}, {Zhao}, \&
  {Kosovichev}}]{Ilonidisetal2011}
{Ilonidis}, S., {Zhao}, J., \& {Kosovichev}, A. 2011, Science, 333, 993

\bibitem[{{Ilonidis} {et~al.}(2012){Ilonidis}, {Zhao}, \&
  {Kosovichev}}]{Ilonidisetal2012}
{Ilonidis}, S., {Zhao}, J., \& {Kosovichev}, A. 2012, Science, 336, 296

\bibitem[{{Jensen} {et~al.}(2001){Jensen}, {Duvall}, {Jacobsen}, \&
  {Christensen-Dalsgaard}}]{Jensenetal2001}
{Jensen}, J.~M., {Duvall}, Jr., T.~L., {Jacobsen}, B.~H., \&
  {Christensen-Dalsgaard}, J. 2001, \apjl, 553, L193

\bibitem[{{Komm} {et~al.}(2009){Komm}, {Howe}, \& {Hill}}]{Kommetal2009}
{Komm}, R., {Howe}, R., \& {Hill}, F. 2009, \solphys, 258, 13

\bibitem[{{Komm} {et~al.}(2008){Komm}, {Morita}, {Howe}, \&
  {Hill}}]{Kommetal2008}
{Komm}, R., {Morita}, S., {Howe}, R., \& {Hill}, F. 2008, \apj, 672, 1254

\bibitem[{{Leka} {et~al.}(2013){Leka}, {Barnes}, {Birch}, {Gonzalez-Hernandez},
  {Dunn}, {Javornik}, \& {Braun}}]{Lekaetal2013}
{Leka}, K.~D., {Barnes}, G., {Birch}, A.~C., {et~al.} 2013, \apj, 762, 130

\bibitem[{{McClintock} \& {Norton}(2016)}]{McClintockNorton2016}
{McClintock}, B.~H. \& {Norton}, A.~A. 2016, \apj, 818, 7

\bibitem[{{Nelson} {et~al.}(2014){Nelson}, {Brown}, {Sacha Brun}, {Miesch}, \&
  {Toomre}}]{Nelson2014}
{Nelson}, N.~J., {Brown}, B.~P., {Sacha Brun}, A., {Miesch}, M.~S., \&
  {Toomre}, J. 2014, \solphys, 289, 441

\bibitem[{{Saidi} {et~al.}(2010){Saidi}, {Burston}, {Moradi}, \&
  {Gizon}}]{Saidi2010}
{Saidi}, Y., {Burston}, R., {Moradi}, H., \& {Gizon}, L. 2010, ArXiv e-prints

\bibitem[{{Scherrer} {et~al.}(2012){Scherrer}, {Schou}, {Bush}, {Kosovichev},
  {Bogart}, {Hoeksema}, {Liu}, {Duvall}, {Zhao}, {Title}, {Schrijver},
  {Tarbell}, \& {Tomczyk}}]{SDO2012}
{Scherrer}, P.~H., {Schou}, J., {Bush}, R.~I., {et~al.} 2012, \solphys, 275,
  207

\bibitem[{{Schunker}(2010)}]{Schunker2010}
{Schunker}, H. 2010, Astronomische Nachrichten, 331, 901

\bibitem[{{Snodgrass} \& {Ulrich}(1990)}]{SnodgrassUlrich1990}
{Snodgrass}, H.~B. \& {Ulrich}, R.~K. 1990, \apj, 351, 309

\bibitem[{{Stein} {et~al.}(2011){Stein}, {Lagerfj{\"a}rd}, {Nordlund}, \&
  {Georgobiani}}]{Steinetal2011}
{Stein}, R.~F., {Lagerfj{\"a}rd}, A., {Nordlund}, {\AA}., \& {Georgobiani}, D.
  2011, \solphys, 268, 271

\bibitem[{{van Driel-Gesztelyi} \& {Green}(2015)}]{vDGGreen2015}
{van Driel-Gesztelyi}, L. \& {Green}, L.~M. 2015, Living Reviews in Solar
  Physics, 12

\bibitem[{{van Driel-Gesztelyi} \& {Petrovay}(1990)}]{vDGP1990}
{van Driel-Gesztelyi}, L. \& {Petrovay}, K. 1990, \solphys, 126, 285

\bibitem[{{Weber} {et~al.}(2013){Weber}, {Fan}, \& {Miesch}}]{Weberetal2013}
{Weber}, M.~A., {Fan}, Y., \& {Miesch}, M.~S. 2013, \solphys, 287, 239

\end{thebibliography}

\onecolumn
\section{List of active regions  included in the SDO/HEAR survey and details of the data reduction.}
{\footnotesize
\begin{longtable}{ l c c c c c | c c | c c }
\caption{Emerging active region and control region tracking locations and emergence time.
The left panel of the table lists the NOAA active region number,   emergence time, Carrington latitude, Carrington longitude, central meridian distance (CMD) at the time of emergence, and the $P$-factor. The middle panel lists the emergence time and  Carrington longitude of the control region. The right panel lists the difference in $B$ angle, $\Delta B = B0\mathrm{(CR)} - B0\mathrm{(EAR)}$, and the rounded difference in days $\Delta T=t_0(\mathrm{CR}) - t_0(\mathrm{EAR})$.\\
The  east-west separation speed of the polarities was not measured for the following active regions: 11074,
11080, 
11098, 
11116, 
11130, 
11143, 
11146, 
11152, 
11156, 
11157, 
11174, 
11194, 
11241, 
11242, 
11267, 
11291, 
11294, 
11311, 
11318, 
11326, 
11334, 
11370, 
11385, 
11396, 
11449, 
11466, 
11560, 
11561, 
11605.\\
{*}{Active regions with a maximum flux larger than the median.}
} \label{tab:ears} \\
 \, \, AR & emergence time &  lat.  & lon. & CMD  & $P$ & CR emergence time & CR lon.  &  $\Delta B0$ &  $\Delta T$ \\
 \, \, \, \#  & [TAI] & [$^\circ$] & [$^\circ$] & [$^\circ$] &  & [TAI] & [$^\circ$] & [$^\circ$] & [days] \\
\hline
\endfirsthead
\multicolumn{3}{l}{{\bfseries \tablename\ \thetable{} -- continued from previous page}} \\
 \, \,  AR & emergence time &  lat.  & lon. & CMD  & $P$ & CR emergence time & CR lon.  &  $\Delta B0$ &  $\Delta T$ \\
 \, \, \, \#  &[TAI] & [$^\circ$] & [$^\circ$] & [$^\circ$] &  & [TAI] & [$^\circ$] & [$^\circ$] & [days] \\
\hline
\endhead
11066   &  2010.05.02\_23:48:00   &    -26.6   &    208.2   &    -16.8   &  0   &  2010.05.10\_23:48:00   &    102.4   &      0.8   &    7 \\ 
11070   &  2010.05.05\_03:24:00   &     20.7   &    195.0   &     -1.5   &  1   &  2010.05.09\_00:00:00   &     89.3   &      0.4   &    3 \\ 
11072*   &  2010.05.20\_17:12:00   &    -15.1   &    314.4   &    -36.1   &  0   &  2010.05.22\_17:12:45   &    288.0   &      0.2   &    2 \\ 
11074   &  2010.05.29\_01:36:00   &     18.6   &    285.4   &     45.3   &  1   &  2010.05.31\_01:36:00   &    258.9   &      0.2   &    1 \\ 
11075   &  2010.05.28\_13:48:00   &    -20.2   &    229.4   &    -17.2   &  1   &  2010.06.11\_13:48:00   &    123.5   &      1.7   &   14 \\ 
11076*   &  2010.05.31\_06:24:00   &    -19.4   &    194.8   &    -16.1   &  0   &  2010.05.20\_06:24:00   &    128.7   &     -1.4   &  -10 \\ 
11079*   &  2010.06.08\_08:24:00   &    -26.0   &    118.5   &     14.5   &  1   &  2010.06.10\_08:23:15   &     92.1   &      0.2   &    1 \\ 
11080*   &  2010.06.10\_02:12:00   &    -23.1   &    109.2   &     28.3   &  2   &  2010.06.12\_02:12:00   &     82.8   &      0.2   &    2 \\ 
11081*   &  2010.06.11\_07:12:00   &     24.0   &    100.5   &     35.6   &  1   &  2010.06.13\_07:12:00   &     74.0   &      0.2   &    1 \\ 
11086   &  2010.07.04\_08:36:00   &     17.8   &    152.0   &     32.2   &  1   &  2010.07.06\_08:36:00   &    125.5   &      0.2   &    1 \\ 
11088   &  2010.07.11\_08:36:00   &    -19.8   &    337.1   &    -50.0   &  0   &  2010.07.13\_08:36:00   &    310.7   &      0.2   &    1 \\ 
11098   &  2010.08.10\_23:12:00   &     13.9   &    300.9   &    -41.2   &  3   &  2010.08.08\_23:12:00   &    327.4   &     -0.1   &   -1 \\ 
11103   &  2010.09.01\_10:12:00   &     26.2   &     85.4   &     26.7   &  4   &  2010.09.10\_00:00:00   &    111.8   &      0.1   &    8 \\ 
11105*   &  2010.09.02\_03:00:00   &     19.0   &     48.5   &     -0.8   &  2   &  2010.09.10\_03:00:00   &     75.0   &      0.0   &    8 \\ 
11114   &  2010.10.14\_04:12:00   &    -20.8   &    246.6   &     32.3   &  0   &  2010.10.06\_04:12:00   &    220.3   &      0.5   &   -7 \\ 
11116   &  2010.10.16\_22:48:00   &     22.5   &    176.8   &     -0.9   &  0   &  2010.10.08\_22:47:15   &    150.4   &      0.6   &   -8 \\ 
11122   &  2010.11.06\_01:12:00   &     13.8   &    261.3   &    -11.3   &  0   &  2010.11.08\_01:12:45   &    235.0   &     -0.2   &    2 \\ 
11130*   &  2010.11.27\_18:12:00   &     13.5   &    327.9   &    -18.7   &  0   &  2010.11.29\_18:12:00   &    301.5   &     -0.2   &    2 \\ 
11132   &  2010.12.03\_23:48:00   &     11.5   &    251.0   &    -13.3   &  3   &  2010.12.05\_23:48:00   &    224.7   &     -0.2   &    1 \\ 
11136   &  2010.12.24\_08:24:00   &    -21.4   &     30.4   &     34.3   &  0   &  2011.01.01\_08:23:15   &     56.7   &     -0.9   &    7 \\ 
11137   &  2010.12.25\_04:48:00   &     17.3   &    309.8   &    -35.1   &  1   &  2010.12.27\_04:48:00   &    283.4   &     -0.2   &    1 \\ 
11138*   &  2010.12.26\_09:36:00   &     12.8   &    318.9   &    -10.3   &  3   &  2010.12.28\_09:36:00   &    292.5   &     -0.2   &    2 \\ 
11141*   &  2010.12.30\_22:36:00   &     34.5   &    267.9   &     -1.4   &  2   &  2010.12.28\_22:36:00   &    294.3   &      0.3   &   -2 \\ 
11142*   &  2010.12.31\_09:24:00   &    -13.8   &    206.0   &    -57.4   &  0   &  2011.01.08\_09:24:00   &    179.7   &     -0.9   &    8 \\ 
11143   &  2011.01.06\_01:12:00   &    -22.1   &    145.6   &    -43.3   &  3   &  2011.01.08\_01:12:45   &    119.2   &     -0.2   &    2 \\ 
11145   &  2011.01.08\_10:00:00   &     15.8   &     97.6   &    -60.1   &  4   &  2011.01.03\_09:59:15   &    163.4   &      0.6   &   -5 \\ 
11146   &  2011.01.10\_11:00:00   &     23.4   &     76.6   &    -54.2   &  1   &  2011.01.15\_10:59:15   &     10.8   &     -0.5   &    4 \\ 
11148   &  2011.01.17\_02:24:00   &    -27.7   &     65.2   &     21.8   &  0   &  2011.01.22\_02:24:45   &    359.3   &     -0.4   &    5 \\ 
11152   &  2011.02.02\_15:36:00   &    -19.1   &    155.9   &    -29.6   &  1   &  2011.02.10\_15:36:45   &    182.2   &     -0.5   &    8 \\ 
11154   &  2011.02.08\_06:00:00   &      7.9   &    151.4   &     39.7   &  1   &  2011.02.10\_06:00:45   &    125.1   &     -0.1   &    2 \\ 
11156   &  2011.02.08\_04:36:00   &    -20.0   &     63.7   &    -48.7   &  2   &  2011.01.31\_04:35:15   &     37.4   &      0.6   &   -8 \\ 
11157*   &  2011.02.08\_01:48:00   &     20.7   &     62.4   &    -51.6   &  4   &  2011.02.10\_01:48:45   &     36.1   &     -0.1   &    2 \\ 
11158*   &  2011.02.11\_01:24:00   &    -19.3   &     35.9   &    -38.8   &  1   &  2011.02.13\_01:23:15   &      9.6   &     -0.1   &    1 \\ 
11159   &  2011.02.10\_06:12:00   &     19.5   &     30.8   &    -54.4   &  4   &  2011.02.12\_06:11:15   &      4.5   &     -0.1   &    1 \\ 
11167   &  2011.03.02\_16:36:00   &     13.5   &    125.6   &    -50.5   &  2   &  2011.02.28\_16:36:00   &    152.0   &      0.0   &   -2 \\ 
11174*   &  2011.03.16\_20:12:00   &     21.3   &     10.7   &     21.0   &  2   &  2011.03.14\_20:12:00   &     37.0   &     -0.0   &   -2 \\ 
11182   &  2011.03.27\_04:12:00   &     13.2   &    201.5   &    -12.0   &  4   &  2011.03.25\_04:12:00   &    227.8   &     -0.1   &   -1 \\ 
11194   &  2011.04.13\_05:12:00   &    -31.8   &      8.9   &     20.3   &  3   &  2011.04.15\_05:11:15   &    342.5   &      0.1   &    1 \\ 
11198   &  2011.04.21\_14:00:00   &    -25.9   &    272.1   &     33.9   &  1   &  2011.04.23\_14:00:45   &    245.7   &      0.2   &    2 \\ 
11199*   &  2011.04.25\_18:36:00   &     21.2   &    187.3   &      4.5   &  2   &  2011.05.09\_18:36:00   &      2.3   &      1.4   &   14 \\ 
11200*   &  2011.04.25\_13:24:00   &    -17.3   &    122.5   &    -63.1   &  4   &  2011.04.27\_13:24:00   &     96.1   &      0.2   &    2 \\ 
11206   &  2011.05.02\_16:48:00   &     23.3   &     91.5   &      0.2   &  3   &  2011.05.13\_16:48:45   &     25.4   &      1.2   &   11 \\ 
11209   &  2011.05.08\_04:48:00   &     34.8   &    358.9   &    -19.6   &  1   &  2011.05.10\_04:48:00   &    332.5   &      0.2   &    1 \\ 
11211   &  2011.05.08\_15:24:00   &    -13.6   &     16.2   &      3.4   &  1   &  2011.05.03\_15:24:00   &     82.3   &     -0.5   &   -4 \\ 
11214*   &  2011.05.13\_18:12:00   &    -24.4   &    275.1   &    -30.0   &  2   &  2011.05.21\_18:12:00   &    169.3   &      0.9   &    8 \\ 
11222   &  2011.05.25\_03:24:00   &     15.8   &    166.9   &     12.5   &  2   &  2011.05.20\_00:00:00   &    272.7   &     -0.6   &   -5 \\ 
11223*   &  2011.05.24\_15:12:00   &    -14.6   &    110.5   &    -50.7   &  4   &  2011.06.01\_15:11:15   &      4.7   &      0.9   &    7 \\ 
11239*   &  2011.06.19\_03:48:00   &     17.0   &    146.0   &    -37.3   &  2   &  2011.06.24\_03:48:00   &     79.9   &      0.5   &    5 \\ 
11241*   &  2011.06.22\_14:24:00   &     19.1   &    108.8   &    -29.0   &  3   &  2011.07.06\_14:24:00   &      2.9   &      1.6   &   14 \\ 
11242*   &  2011.06.28\_10:48:00   &     17.3   &     55.9   &     -4.5   &  1   &  2011.06.26\_10:48:45   &     82.4   &     -0.3   &   -1 \\ 
11267*   &  2011.08.04\_11:48:00   &    -16.4   &    244.4   &    -45.8   &  2   &  2011.07.24\_11:48:00   &    178.3   &     -0.9   &  -10 \\ 
11273   &  2011.08.16\_13:24:00   &    -17.1   &    111.0   &    -19.8   &  2   &  2011.09.08\_13:24:00   &     44.9   &      0.6   &   22 \\ 
11288   &  2011.09.04\_15:12:00   &     18.9   &    183.2   &    -55.5   &  2   &  2011.08.29\_15:12:45   &    209.6   &     -0.1   &   -5 \\ 
11290*   &  2011.09.08\_14:24:00   &    -15.2   &    136.8   &    -49.5   &  3   &  2011.08.28\_14:24:00   &    282.1   &     -0.1   &  -10 \\ 
11291*   &  2011.09.08\_20:36:00   &     23.7   &    158.0   &    -24.9   &  4   &  2011.09.06\_20:35:15   &    184.4   &     -0.0   &   -2 \\ 
11294*   &  2011.09.11\_04:12:00   &    -16.6   &    100.9   &    -51.4   &  0   &  2011.08.28\_04:12:00   &     74.5   &     -0.1   &  -13 \\ 
11297*   &  2011.09.13\_17:48:00   &    -17.6   &    152.3   &     33.9   &  1   &  2011.09.08\_17:48:45   &    218.3   &      0.0   &   -4 \\ 
11300*   &  2011.09.17\_03:48:00   &     24.2   &     92.3   &     19.0   &  0   &  2011.09.24\_00:00:00   &     65.9   &     -0.2   &    6 \\ 
11304   &  2011.09.24\_03:24:00   &     12.4   &    307.4   &    -33.8   &  1   &  2011.09.20\_00:00:00   &    280.9   &      0.1   &   -4 \\ 
11310*   &  2011.10.03\_02:24:00   &    -33.6   &    196.1   &    -26.8   &  1   &  2011.09.25\_02:23:15   &    169.7   &      0.3   &   -8 \\ 
11311*   &  2011.10.03\_16:36:00   &    -12.8   &    177.2   &    -37.9   &  0   &  2011.10.23\_16:36:00   &    273.3   &     -1.4   &   20 \\ 
11318*   &  2011.10.11\_20:12:00   &     20.9   &     94.9   &    -12.6   &  2   &  2011.10.19\_00:00:00   &     68.5   &     -0.5   &    7 \\ 
11322*   &  2011.10.15\_14:24:00   &    -27.0   &    103.5   &     45.5   &  1   &  2011.10.01\_14:24:00   &     37.5   &      0.9   &  -13 \\ 
11326*   &  2011.10.20\_05:12:00   &     15.0   &     21.1   &     24.1   &  3   &  2011.10.22\_05:11:15   &    354.8   &     -0.1   &    1 \\ 
11327*   &  2011.10.19\_08:00:00   &    -21.3   &    334.5   &    -34.3   &  3   &  2011.10.02\_08:00:00   &    198.8   &      1.1   &  -16 \\ 
11331*   &  2011.10.22\_18:36:00   &     10.1   &      5.6   &     42.3   &  1   &  2011.10.20\_18:36:00   &     32.0   &      0.2   &   -2 \\ 
11334*   &  2011.10.30\_00:36:00   &     11.3   &    187.9   &    -39.8   &  2   &  2011.10.28\_00:36:00   &    214.3   &      0.2   &   -2 \\ 
11370   &  2011.12.05\_18:36:00   &    -26.0   &     75.0   &    -28.3   &  3   &  2011.12.07\_18:36:00   &     48.7   &     -0.2   &    1 \\ 
11381*   &  2011.12.17\_10:36:00   &    -19.4   &    260.8   &    -48.8   &  4   &  2011.12.09\_10:36:00   &      6.2   &      1.1   &   -7 \\ 
11385   &  2011.12.22\_04:12:00   &    -30.5   &    225.3   &    -21.9   &  2   &  2011.11.29\_04:12:00   &     80.4   &      3.0   &  -22 \\ 
11396*   &  2012.01.11\_16:12:00   &     25.6   &    287.2   &    -50.0   &  2   &  2012.01.09\_16:12:45   &    313.5   &      0.2   &   -1 \\ 
11397   &  2012.01.12\_22:36:00   &    -20.5   &    277.1   &    -43.3   &  1   &  2012.01.30\_22:36:45   &     92.8   &     -1.6   &   18 \\ 
11400   &  2012.01.14\_02:00:00   &    -13.9   &    295.1   &    -10.3   &  0   &  2012.02.12\_01:59:15   &    150.3   &     -2.2   &   28 \\ 
11404   &  2012.01.15\_07:12:00   &     12.2   &    294.3   &      4.8   &  1   &  2012.01.03\_07:12:45   &    189.0   &      1.3   &  -11 \\ 
11406*   &  2012.01.16\_03:48:00   &    -22.2   &    309.2   &     31.0   &  1   &  2012.02.05\_03:48:00   &    243.4   &     -1.6   &   20 \\ 
11414   &  2012.02.04\_09:24:00   &     -5.4   &     35.7   &     10.8   &  0   &  2012.02.06\_09:23:15   &      9.4   &     -0.1   &    1 \\ 
11416*   &  2012.02.08\_18:24:00   &    -18.5   &    287.6   &    -39.8   &  1   &  2012.02.16\_18:23:15   &    182.2   &     -0.4   &    7 \\ 
11431*   &  2012.03.04\_13:12:00   &    -28.7   &     16.3   &     15.4   &  1   &  2012.03.09\_13:11:15   &    310.5   &      0.0   &    4 \\ 
11437   &  2012.03.16\_16:12:00   &    -34.3   &    167.7   &    -33.4   &  1   &  2012.03.14\_16:12:45   &    194.1   &     -0.0   &   -1 \\ 
11446   &  2012.03.22\_17:24:00   &     24.5   &    103.3   &    -18.1   &  0   &  2012.03.14\_17:23:15   &    208.8   &     -0.2   &   -8 \\ 
11449*   &  2012.03.28\_09:24:00   &    -19.0   &     42.7   &     -4.0   &  3   &  2012.04.06\_09:23:15   &    336.7   &      0.5   &    8 \\ 
11450*   &  2012.03.30\_06:48:00   &     16.0   &    320.6   &    -61.1   &  4   &  2012.04.10\_06:47:15   &    175.5   &      0.6   &   10 \\ 
11456   &  2012.04.11\_07:24:00   &    -20.5   &    217.4   &     -5.6   &  1   &  2012.03.31\_07:24:00   &      2.6   &     -0.7   &  -11 \\ 
11466*   &  2012.04.21\_06:24:00   &     11.7   &     33.1   &    -58.5   &  4   &  2012.04.19\_06:24:00   &     59.5   &     -0.2   &   -1 \\ 
11472*   &  2012.04.29\_05:24:00   &    -28.2   &    294.5   &    -51.9   &  0   &  2012.05.07\_05:24:00   &    268.1   &      0.8   &    8 \\ 
11497*   &  2012.05.31\_13:36:00   &    -22.0   &    223.6   &    -55.1   &  2   &  2012.06.06\_13:35:15   &    197.2   &      0.7   &    5 \\ 
11500   &  2012.06.03\_01:24:00   &     10.0   &    253.4   &      7.8   &  1   &  2012.05.30\_06:21:00   &     39.0   &     -0.5   &   -3 \\ 
11510   &  2012.06.18\_20:36:00   &    -16.2   &     17.8   &    -18.8   &  2   &  2012.06.20\_06:21:00   &    271.9   &      0.1   &    1 \\ 
11511   &  2012.06.21\_11:48:00   &     15.8   &    357.0   &     -4.7   &  1   &  2012.06.10\_11:48:00   &    142.6   &     -1.3   &  -10 \\ 
11523*   &  2012.07.11\_23:24:00   &    -27.4   &     46.3   &    -44.3   &  0   &  2012.07.13\_23:24:00   &     19.8   &      0.2   &    2 \\ 
11531*   &  2012.07.25\_11:12:00   &     14.4   &    308.4   &     36.3   &  2   &  2012.07.30\_11:12:00   &    242.3   &      0.4   &    4 \\ 
11547   &  2012.08.16\_09:36:00   &      5.4   &    297.4   &    -44.7   &  3   &  2012.08.18\_09:36:00   &    270.9   &      0.1   &    1 \\ 
11549   &  2012.08.18\_14:12:00   &    -17.8   &    324.1   &     11.0   &  1   &  2012.08.12\_06:21:00   &    350.5   &     -0.3   &   -6 \\ 
11551   &  2012.08.20\_04:36:00   &     11.9   &    280.8   &    -11.2   &  2   &  2012.08.18\_04:35:15   &    307.2   &     -0.1   &   -2 \\ 
11554*   &  2012.08.23\_07:12:00   &     16.0   &    214.2   &    -36.6   &  3   &  2012.08.25\_07:11:15   &    187.8   &      0.1   &    1 \\ 
11560*   &  2012.08.29\_11:36:00   &      2.9   &    125.4   &    -43.8   &  1   &  2012.08.21\_11:35:15   &    231.1   &     -0.2   &   -8 \\ 
11561   &  2012.08.30\_01:48:00   &    -12.4   &    132.5   &    -28.9   &  1   &  2012.09.10\_01:48:45   &    347.2   &      0.1   &   11 \\ 
11565*   &  2012.09.02\_18:24:00   &      9.8   &     71.8   &    -40.8   &  2   &  2012.09.07\_18:24:45   &      5.8   &      0.0   &    5 \\ 
11570   &  2012.09.11\_19:00:00   &    -12.8   &     10.4   &     16.9   &  0   &  2012.09.13\_18:59:15   &    344.0   &     -0.0   &    1 \\ 
11574*   &  2012.09.16\_13:00:00   &    -24.6   &    301.6   &     10.9   &  0   &  2012.09.14\_13:00:45   &    328.0   &      0.0   &   -1 \\ 
11597*   &  2012.10.17\_19:24:00   &    -21.3   &    251.3   &     13.2   &  2   &  2012.10.15\_19:24:45   &    277.6   &      0.2   &   -1 \\ 
11603   &  2012.10.30\_20:48:00   &      9.4   &     53.2   &    -12.7   &  2   &  2012.10.05\_16:21:00   &    268.2   &      2.0   &  -25 \\ 
11605   &  2012.11.02\_12:24:00   &     17.1   &    333.5   &    -57.4   &  1   &  2012.11.04\_12:24:45   &    307.1   &     -0.2   &    2 \\ 
11607*   &  2012.11.04\_19:48:00   &     12.2   &     30.6   &     30.1   &  1   &  2012.10.30\_19:48:00   &     96.5   &      0.5   &   -4 \\ 
11624   &  2012.11.27\_12:12:00   &     20.7   &     32.5   &    -29.0   &  1   &  2012.11.23\_00:00:00   &    247.5   &      0.6   &   -4 \\ 
\end{longtable}
}

\end{document}